\documentclass{appolb}
\usepackage{graphicx}
\usepackage{amsmath}
\allowdisplaybreaks
\newcommand{\bea}{\begin{eqnarray}} 
\newcommand{\eea}{ \end{eqnarray}} 
\newcommand{\ba}{\begin{eqnarray*}}
\newcommand{\ea}{  \end{eqnarray*}}

\newcommand{\beq}{\begin{equation*} }
\newcommand{\eeq}{\end{equation*}  }

\newcommand{\bqa}{\begin{eqnarray*} }
\newcommand{\eqa}{\end{eqnarray*}}

\begin{document}
\title{Reductions and Contractions of 1-loop \\Tensor Feynman Integrals\vspace*{3mm}
\\
{\it\begin{footnotesize}     Dedicated to the memory of our late 
collegue and friend Jochem Fleischer (1937 - 2013).
The present work is based on a decade of common fruitful research, which would have been impossible without
his impetus, imagination and dedication.\end{footnotesize}}\vspace*{3mm}
%
}
\author{Ievgen Dubovyk\thanks{Presented by I. Dubovyk and T. Riemann at the International Conference
of Theoretical Physics ``Matter To The Deepest'', Ustron 2013}, Janusz Gluza
\address{Institute of Physics, University of Silesia,\\ Uniwersytecka 4, PL-40-007 Katowice, Poland}
\\ \vspace{0.5cm} 
{Andrea A. Almasy, Tord Riemann
}
\address{Deutsches Elektronen-Synchrotron, DESY,\\ Platanenallee 6, 15738 Zeuthen, Germany}
}


\maketitle
\begin{abstract}
 We report on the progress in constructing contracted one-loop tensors. Analytic results for rank $R=4$ tensors, cross-checked numerically, 
 are presented for the first time.  
\end{abstract}
\PACS{11.80.Cr, 12.38.Bx}
  
\section{Introduction}
There are a few source-open programs for 5- and 6-point reductions:  {LoopTools/FF} ($n\le 5, rank\le 4$) --
T. Hahn \cite{Hahn:1998yk,vanOldenborgh:1991yc2},
 {Golem95} -- T. Binoth et al. \cite{Binoth:2008uq}, {PJFry}  ($n\le 5$, rank $\le 5$) -- V. Yundin et al.
\cite{yundin-phd-2012--oai:export,Fleischer:2011zzutphys}. Some of these packages need in addition a 
{library of scalar functions}:
  't Hooft, Veltman
\cite{'tHooft:1978xw},
 {LoopTools/FF},
 {QCDloop/FF} -- K. Ellis and G. Zanderighi \cite{Ellis:2007qk,vanOldenborgh:1991yc2} or
 {OneLOop} with complex masses -- van Hameren \cite{vanHameren:2010cp}.
  In most cases, these packages suffice to calculate one loop processes, however, there are at least two
reasons for improvements. First, which is always desirable, speed improvements. This is important when
calculations are included into precise measurements using Monte Carlo methods. As already discussed  in
\cite{guniaUstron2013}, available methods are at the edge of applications at low energy calculations. Second,
not all of them are able to fulfill high demands concerning accuracy at very specific kinematic points; an
example has been shown in \cite{Gluza:2012yz}.
  
We are working on independent calculations for tensor contractions as an alternative to the PJFry reductions
\cite{pjfry:2011-no-url}, based on work by
Davydychev-Tarasov-Fleischer-Jegerlehner-Riemann-Yundin (DTFJRY)  
\cite{Davydychev:1991vautphys}, \cite{Tarasov:1996br}, \cite{Fleischer:1999hq}, \cite{Fleischer:2010sq},
\cite{Fleischer:2011nt}. First numerical studies have been discussed in \cite{Fleischer:2012adutphys}. In
this material we report on  analytic results for rank 4 and comment on improvements concerning the OLEC
package \cite{olec-project-09utphys}.

\section{Contractions for the 5-point functions with rank $R=4$}

We present here for the first time results of contracted tensors of rank $R=4$
  \bea\label{T54}
I_5^{\mu \nu \lambda \rho}=I_5^{\mu\nu\lambda} \cdot Q_0^{\rho} -\sum_{s=1}^{5} I_4^{\mu\nu\lambda, s} \cdot
Q_s^{\rho},
\eea
for lower rank results, see
\cite{Fleischer:2010sq}, \cite{Fleischer:2011nt}.

 After contraction with chords $q$  (differences of external momenta), we get:
  \bea
q_{a,\mu}q_{b,\nu}q_{c,\lambda}q_{d,\rho}~I_5^{\mu \nu \lambda \rho}
~ =
~ CE_{4,abcd} ~
= ~
-\frac{1}{2} CE_{3,abc} ~ Y_d + C_{5,abcd} .
\eea
Here:
\bea
\label{Q6}
Q_{s}^{\rho}&=&\sum_{i=1}^{5}  q_i^{\rho} \frac{{{s}\choose i}_5}{\left(  \right)_5},~~~ {s}=0,
\ldots , 5.
\eea
The first term $q_{a,\mu}q_{b,\nu}q_{c,\lambda}~I_5^{\mu \nu \lambda}$ is known \cite{Fleischer:2012adutphys},
the second term has to be
determined:
\begin{equation}
 C_{5,abcd} = - \sum_{s=1}^5 q_{a\mu} q_{b\nu} q_{c\lambda} I_4^{\mu\nu\lambda,s}
\frac{1}{2}\left(\delta_{ds}-\delta_{5s}\right),
\end{equation}
and it becomes: 
\begin{eqnarray}
&& C_{5,abcd}  =  \displaystyle\frac{1}{16}\,\, 
\Big\{
	G^5 	
	+\delta_{ab}\delta_{ac} \delta_{ad} G^d 
	-I_1^{5abc}
	-I_1^{5abd}
	-I_1^{5acd}
	-I_1^{5bcd}
	+I_1^{abcd}
 \nonumber \\ 
	& &	-J_3^{a5}
	-J_3^{b5}
	-J_3^{c5}
	-J_3^{d5}
		+\,R^{5ab}
		+\,R^{5ac}
		+\,R^{5bc}
		+\,R^{5da}
		+\,R^{5db}
		+\,R^{5dc}
		 \nonumber \\ 
	& &
		+\,\delta_{bc}\delta_{bd}\left(J_3^{ad}-J_3^{5d}\right)
		+\,\delta_{ac}\delta_{ad}\left(J_3^{bd}-J_3^{5d}\right)
		+\,\delta_{ab}\delta_{ad}\left(J_3^{cd}-J_3^{5d}\right)
	 \nonumber \\ 
	& &	+\,\delta_{ab}\delta_{ac}\left(J_3^{dc}-J_3^{5c}\right)
   		+\,\delta_{ab}\delta_{cd} \tilde{J}_3^{db}
		+\,\delta_{ad}\delta_{bc} \tilde{J}_3^{dc}
		+\,\delta_{ac}\delta_{bd} \tilde{J}_3^{dc}
 \nonumber \\   
   & & 
   		+\,\delta_{ab} \left(\tilde{J}_3^{5b}-R^{b5c}-R^{bd5}+R^{bdc}\right)
		+\,\delta_{ac} \left(\tilde{J}_3^{5c}-R^{c5b}-R^{cd5}+R^{cdb}\right)
\nonumber \\   
   & & 
		+\,\delta_{ad} \left(\tilde{J}_3^{d5}-R^{d5b}-R^{d5c}+R^{dbc}\right)		
		+\,\delta_{bc} \left(\tilde{J}_3^{5c}-R^{c5a}-R^{cd5}+R^{cda}\right)
\nonumber \\   
   & & 
		+\,\delta_{bd} \left(\tilde{J}_3^{d5}-R^{d5a}-R^{d5c}+R^{dac}\right)
		+\,\delta_{cd} \left(\tilde{J}_3^{d5}-R^{d5a}-R^{d5b}+R^{dab}\right)
\nonumber \\   
   & & 
   		+\,\delta_{ab} \delta_{ad} J_4^d Y_c
   		+\,\delta_{ac} \delta_{ad} J_4^d Y_b
		+\,\delta_{bc} \delta_{bd} J_4^d Y_a
   		+\,\delta_{ab} \left(R^{bd}-R^{b5}\right) Y_c
    \nonumber \\ 
	& &		+\,\delta_{ac} \left(R^{cd}-R^{c5}\right) Y_b
   		+\,\delta_{ad} \left(R^{dc}-R^{d5}\right) Y_b
   		+\,\delta_{ad} \left(R^{db}-R^{d5}\right) Y_c
    \nonumber \\ 
	& &		+\,\delta_{bc} \left(R^{cd}-R^{c5}\right) Y_a
   		+\,\delta_{bd} \left(R^{dc}-R^{d5}\right) Y_a
		+\,\delta_{bd} \left(R^{da}-R^{d5}\right) Y_c
 \nonumber \\ 
	& &   		+\,\delta_{cd} \left(R^{da}-R^{d5}\right) Y_b
	 		+\,\delta_{cd} \left(R^{db}-R^{d5}\right) Y_a
   		+\,\left( I_4^d-I_4^5\right) Y_a Y_b Y_c
\nonumber \\
   & & 
   		+\left( I_3^{cd}
   				-I_3^{5c}
   				-I_3^{5d}
   				+R^5
   				+\delta_{cd} R^d
   		\right) Y_a Y_b
    \nonumber \\ 
	& &		+\left( I_3^{bd}
   				-I_3^{5b}
   				-I_3^{5d}
   				+R^5
   				+\delta_{bd} R^d
   		\right) Y_a Y_c
 \nonumber \\
   & & 
   		+\left( I_3^{ad}
   				-I_3^{5a}
   				-I_3^{5d}
   				+R^5
   				+\delta_{ad} R^d
   		\right) Y_b Y_c
  	 \nonumber \\ 
	& &	+\left( I_2^{bcd}
  				-I_2^{5bc}
  				-I_2^{5bd}
  				-I_2^{5cd}
   		  		-J_4^5
   		  		+R^{5b}
   		  		+R^{5c}
   		  		+R^{5d}
   		  \right) Y_a
\nonumber \\
   & & 
   		+\left( I_2^{acd}
   				-I_2^{5ac}
   				-I_2^{5ad}
   				-I_2^{5cd}
   		  		-J_4^5
   		  		+R^{5a}
   		  		+R^{5c}
   		  		+R^{5d}
   		  \right) Y_b
\nonumber \\
   & & 
   		+\left( I_2^{abd}
   				-I_2^{5ab}
   				-I_2^{5ad}
   				-I_2^{5bd}
   		  		-J_4^5
   		  		+R^{5a}
   		  		+R^{5b}
   		  		+R^{5d}
   		  \right) Y_c
\Big\} ,
\end{eqnarray} 
where we have introduced:
\begin{equation}
J_3^{st} \equiv \frac{1}{{s t \choose s t }_5}
\left\{ -{s \choose s }_5 I_3^{[d+],st}+{t s\choose 0 s}_5 R^{ts} - \sum_{u=1}^5 {t s\choose u s}_5 R^{tsu}
\right\},
\end{equation}

\begin{equation}
\tilde{J}_3^{st} \equiv \frac{1}{{s t \choose s t }_5}
\left\{ {s \choose t }_5 I_3^{[d+],st}+{s t\choose 0 t}_5 R^{ts} - \sum_{u=1}^5 {s t\choose u t}_5 R^{tsu}
\right\},
\end{equation} 

\begin{equation}
G^{s} \equiv \frac{1}{{s \choose s }_5}
\left\{ -2 { \choose }_5 R^{[d+],s} + {s\choose 0}_5 J_4^{s} - \sum_{t=1}^5 {s \choose t}_5 J_3^{ts} \right\}.
\end{equation}
$J_4^s$ and $R^{[d+],s}$ are given in Eqs. (2.24) and (2.44)  of \cite{Fleischer:2010sq}, respectively.
For further abbreviations see (2.24), (2.49), (2.9), (2.17), (2.34), (2.41) of \cite{Fleischer:2010sq}.
There also $R^s, R^{st},R^{tsu}$ are defined, $Y_a = Y_{a5} - Y_{55}, Y_{ab} = -(q_a-q_b)^2+m_a^2+m_b^2$. 

Some numerical results for a 5-point function with rank $R=4$ are added
 in \cite{olec-project-09utphys}.
For scalar functions we use the OneLOop package and compare the results with LoopTools/FF. 
For the considered kinematic points full agreement has been obtained.
The kinematics is that of the process $e^-e^+\to \mu^-\mu^+ \gamma$.
Using an MC generator, we have checked thousands of points up to rank
three. All of them agreed between OLEC and LT/OneLOop.
 For ranks 3 and the new results presented here we
made checks also against the  public version of Golem95, available at
\verb+http://golem.hepforge.org/95/+. At the OLEC webpage \cite{olec-project-09utphys}, we added two sets of
files with the
output. In set I, the kinematics is chosen such that 3-point functions hit the IR
singularities, while in set II the 3-point functions are slightly off the IR
singularities. The Golem95 results were different from LT,
OneLOop and OLEC, starting already from rank 3.\footnote{The problem in Golem95 v.1.2.1
has been settled in the meantime for the set I with changelog 128 (11 Oct 2013),
{\tt https://golem.hepforge.org/trac/changeset/128}.}
For rank four we give below such an example:
\begin{verbatim}
p1s =  1.1163688400000000E-002  p2s =  2.6109999999999998E-007  
p3s =  0.0000000000000000       p4s =  2.6109999999999998E-007  
p5s =  1.1163688400000000E-002
s12 = -0.70858278190000001      s23 = -1.5343299000000002E-003  
s34 = -0.12851860429999998      s45 = -0.61023937949999996      
s15 =  0.92668942420000000      m1s =  1.1163688361676107E-002  
m2s =  0.0000000000000000       m3s =  2.6112003932088364E-007 
m4s =  2.6112003932088364E-007  m5s =  0.0000000000000000 
\end{verbatim}

\begin{verbatim}
The R=4 contractions, a,b,c,d=3,3,3,3
OLEC:   ( -48094.1074 54542318     , -47802.08746 5035322 )
LoopTools: ( -48094.1074 65           , -47802.08746 05      )
The R=4 contractions, a,b,c,d=3,3,3,4  
OLEC:   ( -18463.1204 24842149     , -23446.4704 12257226 )
LoopTools: ( -18463.1204 31           , -23446.4704 09       )
The R=4 contractions, a,b,c,d=3,3,3,5  
OLEC:   (  0.0000000000000000     ,  0.0000000000000000 )
LoopTools  (  0.0000000000000000     ,  0.0000000000000000 )
\end{verbatim}
The last result with d=5 is a virtue of the construction of chords where $q_5=0$.  
\section{OLEC package, Fortran code} 
The idea of external contractions has been implemented for the first time in the C++ code OLEC 
  for tensors up to rank 3 \cite{Fleischer:2012adutphys}, and basic examples are given at
\cite{olec-project-09utphys}. In the meantime, a Fortran code has been written with the aim of further
optimization.
The calculation of contracted tensor integrals consists of two basic steps. The first and most time consuming 
step is a preparation
of building blocks for a calculation - basic scalar integrals $(\sim 50\%)$ and so-called signed minors $(\sim
25\%)$. 
 We cannot make much about scalar integral libraries, unless new independent developments appear. For some
recent efforts in this direction, see  \cite{Guillet:2013mta}. 
However, the {\it calculation} of signed minors can be improved, both the cache system and and the
computation algorithm. 
The second step is {\it performing} of the contractions $(\sim 25\%)$, which also can be improved. 
During the initialization procedure scalar integrals and   signed minors are calculated and stored in RAM
memory.
For efficient usage of memory  appropriate data structures are constructed. E.g. for 5-point kinematics we
have in total only 30 different
basic scalar integrals  while for signed minors situation is more complicated because all minors with up to 4 scratched rows and columns  
are needed. The ``natural'' storage model for such objects in the form of multidimensional arrays is not
an optimal solution. For example, for minors of rank 3 (3 rows and columns are excluded) one has an array with 46656 elements
but according to the minors' symmetry properties only 210 of them are different and non zero. For the purpose
of minimal
memory usage a linear storage model is chosen. In general minors of a given rank are stored in the memory as one--dimensional arrays
and access to them can be done according to the following pattern:
$minor_{rank}[AddressTable[func(i,j,k...)]]$, as it was first done in \cite{pjfry:2011-no-url}. For a given
minor indices 
$i,j,k,...$ a function $func$ based on bitwise operations calculates an address of an element from the constant 
table $AddressTable$. This table is the same for all minors and contains positions of values of minors in array $minor_{rank}$.
Finally the cache contains a few hundred of double precision numbers for all building blocks like minors, scalar integrals and auxiliary functions. 
\\
In the new version of the library the computation algorithm for signed minors has also been changed. In practical computations
  of one-loop cross--sections usually at least  all contractions up to rank 3
are needed (like in QED). It means that minors of all ranks are needed as well. The procedure is iterative, cache is filled first from minors of the highest rank 4
(with 8 indices) which are used further to calculate minors of rank 3. And so on.
All loops connected with minor indices during this procedure are unrolled and the highest level of optimization is applied for compilation (these are algebraical manipulations for which -O3 
optimization does not spoil double accuracy). Unrolling and usage of optimalization for algebraic part of the package  allows to decrease  extremely the matrix algebra computation time.

\section{Summary and Outlook}
In recent years, the strategy described has been developed in a bunch of papers for 
explicit  {analytical and recursive treatment} of {  heptagon, hexagon and pentagon tensor
integrals} of rank 
$R$ in terms of
pentagons and boxes of rank $R-1$. 
A systematic derivation of expressions which are explicitly  {free of inverse Gram determinants}   {$()_5$}
until pentagons of rank $R=5$
has been worked out.
The numerical   {package OLEC for contracted tensor integrals in C++ and
Fortran \cite{olec-project-09utphys,AFGRprep} are under development and tests.

Our preliminary benchmarks show that the OLEC library
in the present form is faster than 5-point tensor reductions implemented in LoopTools/FF by about an order of
magnitude (tested up to rank 3). 
Work in progress includes
 a correct treatment of small Gram determinant cases for the reductions (expansion in small parameters
\cite{pjfry:2011-no-url} or 
using hypergeometric representations \cite{Fleischer:2003rm}), adding rank 4 (this paper) and rank 5
contractions for 5-point functions, programming contracted tensors for 6- and 7- point functions.  
 
\section*{Acknowledgments}

Work supported by European Initial Training Network LHCPHENOnet PITN-GA-2010-264564.  



\providecommand{\href}[2]{#2}

\end{document}